\documentclass[usenatbib]{mn2e}
\usepackage{graphicx, color, pdflscape}

\newcommand{\etal}{{\it et al.}}
\newcommand{\ie}{{\it i.e.}}

\newcommand{\be}{\begin{equation}}
\newcommand{\ee}{\end{equation}}

\usepackage{epsfig}
\usepackage{color}
\begin{document}

\title[Galaxy Zoo: Unwinding the Winding Problem]{Galaxy Zoo: Unwinding the Winding Problem -- Observations of Spiral Bulge Prominence and Arm Pitch Angles Suggest Local Spiral Galaxies are Winding}
\author[K.L. Masters \etal]{Karen L. Masters$^{1,2}$, Chris J. Lintott$^{3}$, Ross E. Hart$^{4}$, Sandor J. Kruk$^{3,4}$, \newauthor Rebecca J. Smethurst$^{5}$,  Kevin V. Casteels$^6$, William C. Keel$^7$, Brooke D. Simmons$^{8,9}$, \newauthor Dennis Olivan Stanescu$^{10}$, Jean Tate$^{10}$, Satoshi Tomi$^{10}$\\
$^1$Haverford College, Department of Physics and Astronomy, 370 Lancaster Avenue, Haverford, Pennsylvania 19041, USA\\
$^2$Institute for Cosmology and Gravitation, University of Portsmouth, Dennis Sciama Building, Burnaby Road, Portsmouth, PO1 3FX, UK \\
 $^{3}$Oxford Astrophysics, Department of Physics, University of Oxford, Denys Wilkinson Building, Keble Road, Oxford, OX1 3RH, UK\\
 $^{4}$European Space Agency, ESTEC, Keplerlaan 1, NL-2201 AZ, Noordwijk, The Netherlands\\
 $^5$Centre for Astronomy \& Particle Theory, University of Nottingham, University Park, Nottingham, NG7 2RD, UK\\
 $^6$Department of Physics and Astronomy, University of Victoria, Victoria, BC V8P 1A1, Canada\\
 $^7$Department of Physics \& Astronomy, 206 Gallalee Hall, 514 University Blvd., University of Alabama, Tuscaloosa, AL 35487-0234, USA\\
 $^8$Physics Department, Lancaster University, Lancaster, LA1 4YB, UK\\
 $^9$Center for Astrophysics and Space Sciences (CASS), Department of Physics, University of California, San Diego, CA 92093, USA\\
 $^{10}$Galaxy Zoo Citizen Scientist\\
\\
 $^*$This publication has been made possible by the participation of hundreds of thousands of volunteers in the Galaxy Zoo project. \\ Their contributions are individually acknowledged at \texttt{http://authors.galaxyzoo.org}. \\
\\
{\tt E-mail: klmasters@haverford.edu}
 }

\date{Accepted 2019 April 23. Received 2019 April 5; in original form 2018 August 11}
\pagerange{1--12} \pubyear{2019}

\maketitle

\begin{abstract}

We use classifications provided by citizen scientists though Galaxy Zoo to investigate the correlation between bulge size and arm winding in spiral galaxies. Whilst the traditional spiral sequence is based on a combination of both measures, and is supposed to favour arm winding where disagreement exists, we demonstrate that, in modern usage, the spiral classifications Sa--Sd are predominantly based on bulge size, with no reference to spiral arms. Furthermore, in a volume limited sample of galaxies with both automated and visual measures of bulge prominence and spiral arm tightness, there is at best a weak correlation between the two. Galaxies with small bulges have a wide range of arm winding, while those with larger bulges favour tighter arms. This observation, interpreted as revealing a variable winding speed as a function of bulge size, may be providing evidence that the majority of spiral arms are not static density waves, but rather wind-up over time. This suggests the ``winding problem" could be solved by the constant reforming of spiral arms, rather than needing a static density wave. We further observe that galaxies exhibiting strong bars tend have more loosely wound arms at a given bulge size than unbarred spirals. This observations suggests that the presence of a bar may slow the winding speed of spirals, and may also drive other processes (such as density waves) which generate spiral arms. It is remarkable that after over 170 years of observations of spiral arms in galaxies our understanding of them remains incomplete.
\end{abstract}


\section{Introduction}

The classification of objects into categories is a common technique in many areas of science. Galaxy morphology (the shapes and features seen in images of galaxies) was the most obvious starting point for this process in extragalactic astronomy. As a result many galaxy classification schemes have been developed (see \citealt{Buta2013}, and \citealt{Sandage2005}\footnote{In which can also be found instructions for simulating the structures seen in galaxies using cream in coffee or frozen butter sticks in milk} for recent reviews).
The scheme first laid out by \citet{Hubble1926,Hubble1936}, and used in revised and expanded versions such as {\it The Hubble Atlas} by \citet{Sandage1961}, the {\it Third Reference Catalogue of Bright Galaxies}, or RC3 by \citet{1991rc3..book.....D} remains the basis of the most commonly used classifications. 

The basic ``Hubble sequence" splits galaxies into ``spiral" and ``elliptical" types, labelling ellipticals by their degree of elongation (from E0 being completely round, to E7 ellipticals the most ``cigar-like''). Spiral galaxies are then ordered in a sequence extending away from the ellipticals, split into two branches by the presence or absence of a galactic bar. 
\citet{Hubble1936} also correctly discussed the existence of an intermediate type (lenticulars, or S0s), even though no examples were known at the time \citep{Buta2013}. 

\begin{landscape}
\begin{figure}
\includegraphics[width=160mm,angle=-90]{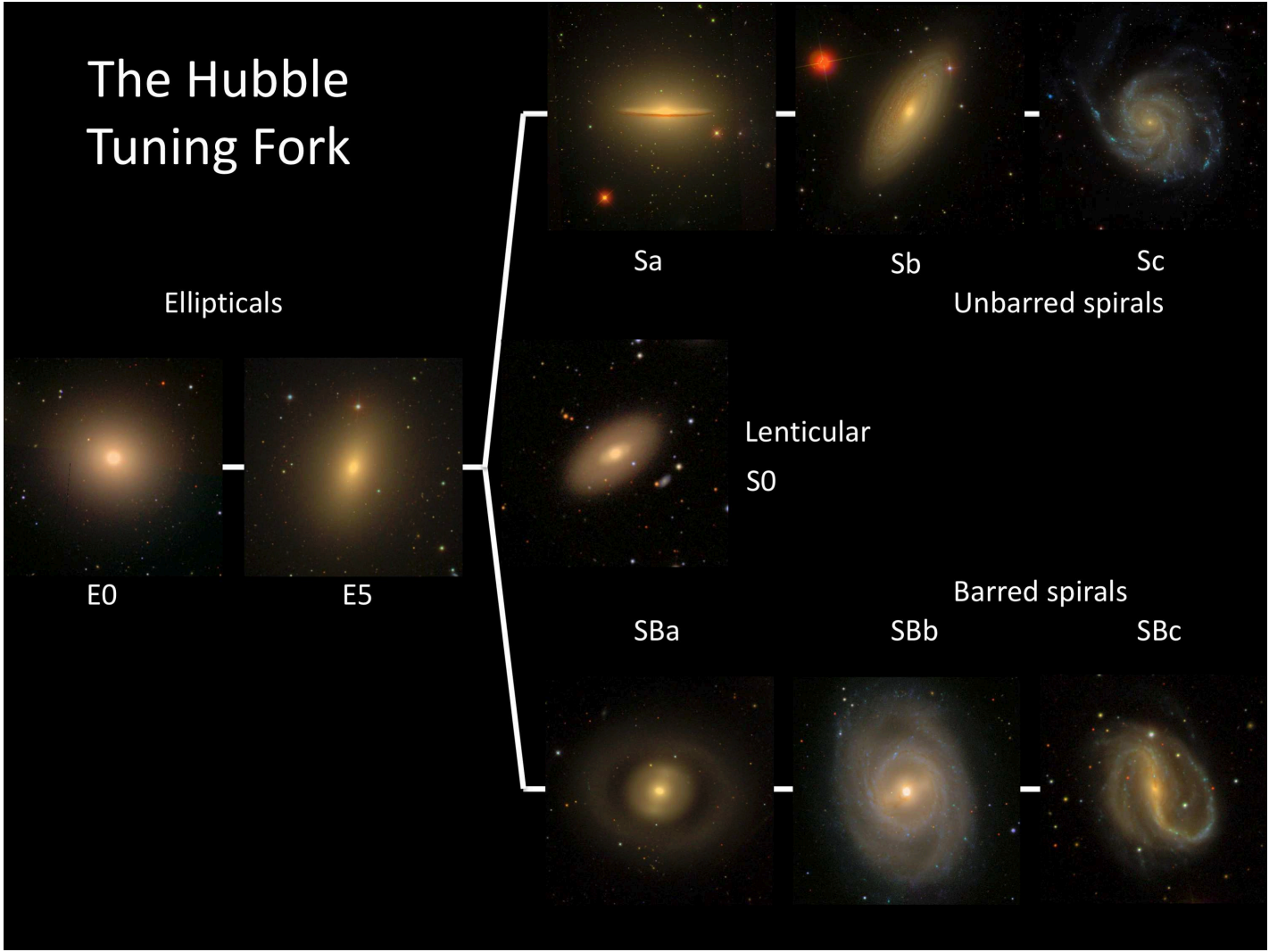}
\caption{The Hubble Sequence illustrated by the examples suggested by \citet{Hubble1926} with images from the Sloan Digital Sky Survey \citep[SDSS; ][]{York2000}. The galaxies are: E0 -- NGC 3379 (M105); E5 -- NGC 4621 (M59);  Sa -- NGC 4594 (The Sombrero);  Sb -- NGC 2841; Sc -- NGC 5457 (M101 or The Pinwheel); SBa -- NGC 2859; SBb -- NGC 3351 (M95); SBc -- NCG 7479. We have also included an S0 (NGC 6278); only theorised in Hubble's original scheme as no examples were known at the time. \label{sequence}}
\end{figure}
\end{landscape}

The original Hubble sequence of Sa-Sb-Sc spiral galaxies (\citealt{Hubble1926}, extended to Sd by \citealt{1959HDP....53..275D}) was set up using three distinct criteria. These were based on (1) spiral arm appearance, split into (a) how tightly wound the spiral arms are and (b) how clear, or distinct, the arms are, and (2) the prominence of the central bulge. Sa galaxies were described as having large bulges and tight, smooth (very distinct) arms, while in contrast typical Sc galaxies were described as having very small ``inconspicuous" bulges and very loose patchy (indistinct) arms. In Hubble's language ``normal" (S) and ``barred" (SB) spirals had identical parallel sequences. These types are illustrated in Figure \ref{sequence} by the example galaxies given in \citet{Hubble1926}. 

By analogy with the terminology used for stellar classification (and explicitly making the point that this was not a comment on the expected evolution of galaxies\footnote{See the Footnote I on page 326 of \citet{Hubble1926}, and also \citet{Baldry2008AG}}), Hubble dubbed the spiral types (a) ``early",  (b) ``intermediate" and (c) ``late"-type. This was the basis of sometimes confusing terminology which has stuck, with astronomers now more commonly using ``early-type galaxies"  (ETG) to refer to elliptical and lenticular galaxies (often, but not always, excluding the ``early-type" or Sa spirals, e.g. as used by the ATLAS-3D team; \citealt{2011MNRAS.413..813C,2011MNRAS.416.1680C}; or the SAMI team \citealt{2018arXiv180711547F}; and also see \citealt{Stanford1998} for an earlier example); while ``late-type" is commonly used to refer to any spiral galaxy (but sometimes excludes Sa spirals, e.g. \citealt{Strateva2001}).

The morphology of a galaxy encodes information about its formation history and evolution through what it reveals about the orbits of the stars in the galaxy, and is known to correlate remarkably well with other physical properties \citep[e.g. star formation rate, gas content, stellar mass][]{RobertsHaynes1994, Kennicutt1998, Strateva2001}. These correlations, along with the ease of automated measurement of colour or spectral type, have resulted in a tendency for astronomers to make use of classification on the basis of these properties rather than morphology {\it per se} (to select just a few examples\footnote{With thanks to the participants of the ``Galaxy Zoo Literature Search" for finding many of these}: \citealt{Bell2004, Weinmann2006, vandenBosch2008, Cooper2010, Zehavi2011}). Indeed the strength of the correlation has led some authors to claim that the correspondence between colour and morphology is so good that that classification by colour alone can be used to replace morphology \citep[e.g.][]{ParkChoi2005, Faber2007, AscasibarAlmeida2011}, or to simply conflate the two (e.g. \citealt{TalvanDokkum2011}; but see \citealt{vandenBergh2007} for a contrary view). Meanwhile the size of modern data sets (e.g. the Main Galaxy Sample, MGS of the Sloan Digital Sky Survey, SDSS, \citealt{Strauss2002}) made the traditional techniques of morphological classification by small numbers of experts implausible. This problem was solved making use of the technique of crowdsourcing by the Galaxy Zoo project \citep{Lintott2008,Lintott2011}. One of the first results from the Galaxy Zoo morphological classifications was to demonstrate on a firm statistical basis that colour and morphology are not equivalent for all galaxies (as first presented in \citealt{Bamford2009, Schawinski2009,Masters2010}), making it clear that morphology provides complementary information to stellar populations (traced by either photometry or spectra) to understand the population of galaxies in our Universe.   

 In this article we explore an updated view of the Hubble spiral sequence obtained from visual classifications provided by 160,000 members of the public on $\sim$ 250,000 galaxies from the SDSS MGS \citep{Strauss2002}. These classifications are described in detail in \citet{Willett2013}, available to download from {\tt data.galaxyzoo.org} (as well as being included in SkyServer.org as an SDSS Value Added Catalogue from DR10; \citealt{DR10}). The basic division into spiral--elliptical (or featured--smooth in the language of Galaxy Zoo, which corresponds to what many astronomers mean by late- and early-type) galaxies has been discussed at length \citep[e.g.][]{Willett2013}. In this article we particularly focus on the spiral (or more precisely ``featured, but not irregular") sequence, and investigate whether the traditional criteria for the ordering of spiral galaxies along this sequence fit in with the picture revealed by Galaxy Zoo morphologies. 

\begin{figure*}
\includegraphics[height=160mm,angle=90]{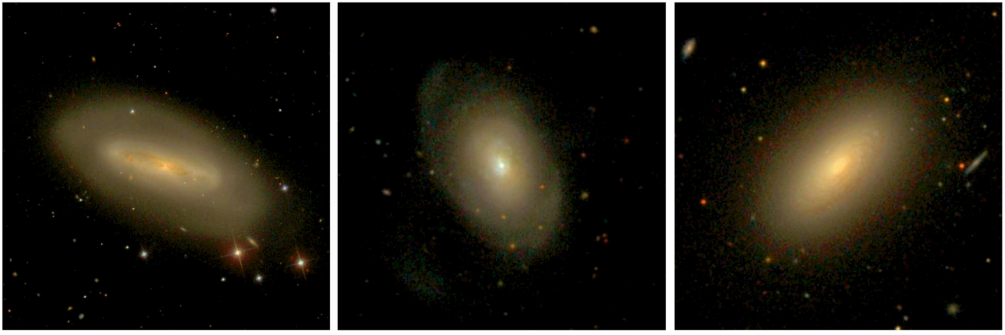}
\caption{Examples of Sa galaxies with large, intermediate and small bulges from the classifications by \citet{Hogg1993}.  The galaxies are (from left to right) large bulge Sa: NGC 2639; intermediate bulge Sa: NGC 3611; small bulge Sa: NGC 4293. All images are $gri$ composites from SDSS. \label{Sa}}
\end{figure*}

Among experts in morphology \citep[e.g.][]{Sandage2005,Buta2013}, there has been a consensus that for most spiral galaxies the traditional criteria involving both spiral arm appearance and bulge size result in consistent classification. \citet{Buta2013} explains, however, that ``in conflicting cases, emphasis is usually placed on the appearance of the arms". Examples of conflicting cases, particularly of galaxies with tightly wound spirals and small bulges can be easily  found in the literature (e.g. examples from \citealt{Hogg1993} are shown in Figure \ref{Sa}; also see \citealt{Sandage1961, SandageBedke1994, Jore1996}), and the existence of ``small bulge Sa galaxies" (as defined by their arm types) had been recognised even in Hubble's time (according to \citealt{Sandage2005}). \citet{Buta2013} also explains that SB (strongly barred spiral) galaxies with small bulges may commonly have tightly wound arms, and therefore be classed as SBa. This traditional picture of the spiral sequence is best illustrated in Figure 7 of \citet{kennicutt1981} which shows just how strongly measurements of pitch angle correlate with traditional determinations of Hubble type from \citet{sandagetammann1981}. 

However it is also clear that modern automatic galaxy classification has tended to conflate bulge size alone with spiral type \citep[e.g.][]{Goto2003, Laurikainen2007, gadotti2009, Masters2010, Lange2016}. Furthermore, automatic classification of galaxies into ``early-" and ``late-" types (\ie ~referring to their location on the Hubble Sequence) based on bulge-total luminosity ratio ($B/T$) or some proxy for this through a measure of central concentration, or light profile shape (e.g. Sersic index, as reviewed by \citealt{2005PASA...22..118G}), has become common \citep[e.g.][]{vanderWel2011,Wang2019}. Indeed, \citet{Sandage2005} reveals this is not new, claiming ``the Hubble system for disk galaxies had its roots in an arrangement of spirals in a continuous sequence of decreasing bulge size and increasing presence of ``condensations'' over the face of the image that had been devised by \citet{Reynolds1920}", and explaining that efforts to classify galaxies on the basis of the concentration of their light alone were first begun by \citet{Shapley1927}. 

 For example, while the current classification of a lenticular (or S0) galaxy usually assumes a dominent bulge component, early S0 classification included galaxies with bulges of different sizes (S0a-S0c; \citealt{SpitzerBaade1951, vandenBergh1976}), a classification recently revived by ATLAS-3D in their morphology ``comb" which includes parallel sequences of star forming and passive (or anaemic) spirals, and an ETG fast-rotator bulge size sequence similar to the S0 sequence \citep{2011MNRAS.416.1680C}, as well as by \citet{Kormendy2012} in their parallel lenticular classification scheme explicitly based on $B/T$. The lenticular classification is therefore an extreme example of the same name being used to represent many different classifications of galaxies. 

 Even within spiral galaxies, it has been understood for some time that the diversity of spiral arms observed in galaxies is not perfectly captured by the Sa-Sb-Sc spiral arm descriptors. As discussed at length by \citet{Buta2013}, the number of arms (commonly denoted $m$), ``character" of the arms (e.g. ``grand-design" or ``flocculent" as first described by \citealt{Elmegreen1981}) and the sense of the winding of the arms relative to the galaxy rotation are all additional dimensions which can be used for classification (also see \citealt{1987ApJ...314....3E,AnnLee2013}). \citet{Buta2013} notes that most low $m$ spirals are grand design in character, and goes on to discuss how spiral arm ``character" is thought to link to typical formation mechanism (with grand design spirals linked to density wave mechanisms, and flocculent spirals suggested to come from sheared self-propagating star formation regions). While laying out the distinction between spiral types, \citet{Elmegreen1981} also note that the differences suggested different spiral arm formation mechanisms. Early analytic models for spiral arms described density waves \citep{LinShu1964}, while the first simulations of spiral structure in isolated galaxies resulted in much more flocculent types (e.g. see the review in \citealt{Elmegreen1981}, and in  \citet{DobbsBaba2014} you will find a comprehensive recent review of more recent simulations of spiral structure). Even in recent simulations, two-armed ``grand design" spirals are difficult to form, without some kind of strong perturber \citep{Sellwood2011}. Recent observational work using Galaxy Zoo identifications of arm number \citep{Hart2016,Hart2017a} found that two-armed (``grand design") spirals are redder in colour than those with many arms (\ie ~flocculent spirals), providing more evidence of a link between arm formation mechanisms and star formation properties.  

Spiral bulges have also been revealed to have diversity -- with a distinction needing to be made between ``classical" bulges (spheroidal and pressure supported systems with an $R^{1/4}$ or Sersic $n=4$ profile) and ``pseudo" or ``discy" bulges (which are rotationally supported and having an exponential, or Sersic $n=1$ profile; \citealt{gadotti2009,kormendykennicutt2004}). It is observed that the stellar populations of these two types of bulges are noticeably different \citep{FisherDrory}, and it is generally assumed that the former is formed in galaxy merging, while the latter could be grown via secular evolution driving radial flows \citep[e.g.][]{gadotti2009}. 

In this paper we make use of Galaxy Zoo classifications, which provide a quantitative visual description of structures seen in local galaxies, capturing the typical range of descriptions used to construct the traditional Hubble sequence, but are not tied to any specific classification scheme (e.g. a spiral galaxy might easily be described has having tightly wound spiral arms and a large bulge, if this is how it looks). We review these classifications in Section \ref{sample}, give basic demographics of the local sample in Section \ref{demographics}, and discuss how to use them to construct a traditional Hubble sequence, along with the implications of trends of various visible structures in Section \ref{discussion}. We conclude with a summary section. Where distances are needed a value of $H_0 = 70$km/s/Mpc is used; the galaxies are sufficiently nearby that other cosmological parameter choices make a negligible difference. 

\section{Sample and Data} \label{sample}

The first two phases of Galaxy Zoo (which ran from July 2007-April 2010\footnote{GZ1 is archived at {\tt http://zoo1.galaxyzoo.org}, and GZ2 at {\tt http://zoo2.galaxyzoo.org}}) were entirely based on imaging from the Legacy Survey of the Sloan Digital Sky Survey (SDSS; \citealt{York2000}). In this paper we make use exclusively of classifications from the second phase of Galaxy Zoo (or GZ2; \citealt{Willett2013}). In total, almost 300,000 images of galaxies were shown in GZ2, selected to represent the largest (in angular size) and brightest galaxies observed by SDSS. For full details of the sample selection see \citet{Willett2013}, but in brief GZ2 made use of the SDSS DR7 imaging reduction \citep{DR7} and selected galaxies with r-band apparant magnitude, $m_r<17.0$, radius $r_{90}>3\arcsec$ (where $r_{90}$ is the radius containing 90\% of the r-band Petrosian aperture flux) and $0.0005<z<0.25$. Image cutouts were generated as $gri$ colour composites centred on each galaxy with a size $8.48r_{90}\arcsec \times 8.48r_{90}\arcsec$. 

Visual classifications for GZ2 were collected via a web interface, which presented volunteers with the colour cutout, and a selection of simple questions about the object shown. Following \citet{Willett2013} (hereafter W13) we define a {\it classification} as the sum of all information provided about a galaxy by a single user. These {\it classifications} are made up of answers to a series of {\it tasks} presented in a decision tree. A flow chart of this tree is presented as Figure 1 in W13, and for the convenience of the reader we reproduce Table 2 of W13 which summarises all possible {\it tasks} and answers in our Table \ref{tbl-tree}. 

\begin{table}
 \caption{The GZ2 decision tree, comprising 11 tasks and 37 responses. The `Task' number is an abbreviation only and does {\em not} necessarily represent the order of the task within the decision tree. The text in `Question' and `Responses' are displayed to volunteers during classification. `Next' gives the subsequent task for the chosen response. \label{tbl-tree}}
 \begin{tabular}{@{}cllr}
 \hline
\multicolumn{1}{l}{Task} &
\multicolumn{1}{c}{Question} &
\multicolumn{1}{c}{Responses} &
\multicolumn{1}{c}{Next} 
\\ 
\hline
\hline						
01    & {\it Is the galaxy simply smooth   }  & smooth           & 07 \\
      & {\it and rounded, with no sign of  }  & features or disk & 02 \\
      & {\it a disk?                       }  & star or artifact & {\bf end} \\
      \hline
02    & {\it Could this be a disk viewed   }  & yes              & 09 \\
      & {\it edge-on?                      }  & no               & 03 \\
      \hline
03    & {\it Is there a sign of a bar      }  & yes              & 04 \\
      & {\it feature through the centre    }  & no               & 04 \\
      & {\it of the galaxy?                }                                        \\
      \hline
04    & {\it Is there any sign of a        }  & yes              & 10 \\
      & {\it spiral arm pattern?           }  & no               & 05 \\
      \hline
05    & {\it How prominent is the          }  & no bulge         & 06 \\
      & {\it central bulge, compared       }  & just noticeable  & 06 \\
      & {\it with the rest of the galaxy?  }  & obvious          & 06 \\
      & {\it                               }  & dominant         & 06 \\
      \hline
06    & {\it Is there anything odd?        }  & yes              & 08 \\ 
      & {\it                               }  & no               & {\bf end}        \\
      \hline
07    & {\it How rounded is it?            }  & completely round & 06 \\
      & {\it                               }  & in between       & 06 \\
      & {\it                               }  & cigar-shaped     & 06 \\
      \hline
08    & {\it Is the odd feature a ring,    }  & ring             & {\bf end}        \\
      & {\it or is the galaxy disturbed    }  & lens or arc      & {\bf end}        \\
      & {\it or irregular?                 }  & disturbed        & {\bf end}        \\
      & {\it                               }  & irregular        & {\bf end}        \\  
      & {\it                               }  & other            & {\bf end}        \\  
      & {\it                               }  & merger           & {\bf end}        \\  
      & {\it                               }  & dust lane        & {\bf end}        \\  
      \hline
09    & {\it Does the galaxy have a        }  & rounded          & 06 \\
      & {\it bulge at its centre? If       }  & boxy             & 06 \\
      & {\it so, what shape?               }  & no bulge         & 06 \\
      \hline
10    & {\it How tightly wound do the      }  & tight            & 11 \\
      & {\it spiral arms appear?           }  & medium           & 11 \\
      & {\it                               }  & loose            & 11 \\    
      \hline
11    & {\it How many spiral arms          }  & 1                & 05 \\
      & {\it  are there?                   }  & 2                & 05 \\
      & {\it                               }  & 3                & 05 \\
      & {\it                               }  & 4                & 05 \\
      & {\it                               }  & more than four   & 05 \\
      & {\it                               }  & can't tell       & 05 \\
\hline
 \end{tabular}
\end{table}

Each galaxy was classified by $\sim$40 volunteers, and their inputs combined via what we call ``consensus algorithms". W13 describes in detail the process by which user responses are weighted and combined to provide vote fractions for each answer to each task for each galaxy in GZ2. We will refer to vote fractions as $p_{\rm xxx}$, where ``xxx" will describe the answer of interest. For example $p_{\rm features}$ will refer to the fraction of users answering {\it task} 01 by indicating they could see ``features or a disc" in the galaxy. W13 also describes a process of correcting for classification bias, caused primarily by galaxies at larger redshift appearing dimmer and at coarser physical resolution than if viewed at lower redshift. \citet[hereafter H16]{Hart2016} investigate this classification bias further, especially with regard to the visibility of spiral arms in GZ2, and update the redshift debiasing method to provide an improved set of debiased classifications from GZ2. In this paper we make use of the debiased classifications from H16, and when we use the terminology $p_{\rm xxx}$ we specifically refer to the debiased vote fraction using the H16 debiasing.

We select a low redshift volume limited sample, which is similar to the sample selection of H16 (and \citealt{Hart2017a}). This is motivated by the desire to have galaxies with sufficient angular resolution that spiral arm features can be clearly identified as well as to limit the impact of redshift debiasing.  Of the galaxies in GZ2 \citep{DR7,Strauss2002}, we select the 22,045 which have measured redshifts in the range $0.01<z<0.035$, and which have an $r$-band absolute Petrosian magnitude (de-reddened and k-corrected to $z=0$ following \citealt{Bamford2009}) of $M_r < -19.0$. The $r$-band imaging from the Legacy Survey programme of the Sloan Digital Sky Survey (SDSS \citealt{York2000}), has a mean FWHM seeing of 1.2\arcsec ~\citep{Kruk2018}\footnote{The commonly cited value of 1.4\arcsec ~for median SDSS seeing is an overestimate of the final quantity, largely because it was based only the early data release (EDR) imaging -  for the 
 footprint, the best seeing imaging was kept in areas which had repeat visits \citep{Ross2011}; see also {\tt https://www.sdss.org/dr15/imaging/other\_info/}}, which provides a physical resolution of 0.1-0.8 kpc at the redshifts of this sample, enabling the reliable visual identification of small bulges, and spiral arms. 

 We remove {six} galaxies which have more than 50\% of their classification votes for ``star or artifact". Inspecting these objects reveals that they are typically genuine galaxies, but with corrupted images (e.g. under a satellite trail, or diffraction spike from a nearby bright star). However, we are not able to construct a useful GZ2 consensus classification since so many people marked them as artifacts.

In addition to identifying the spiral galaxies of interest for this work, an identification of ``features" in a galaxy via the Galaxy Zoo method might indicate disturbed or irregular morphology or mergers (or any other observed features, e.g. dust lanes). Users could identify these in GZ2 after indicating the that the galaxy showed ``odd" features, and then indicating what they thought was odd. All users classifying a galaxy answered the question ``Is there anything odd?". We select for disturbed, irregular or merging galaxies by requiring that $p_{\rm odd} > 0.42$ and $N_{\rm classifier}>20$ (as recommended in W13), and further selecting galaxies for which $(p_{\rm irregular}+p_{\rm disturbed} + p_{\rm merger} > 0.6$ (\ie ~approximately 60\% or more of the classifiers who indicated the galaxy was ``odd" thought the reason was that it was either irregular, disturbed or merging). As users could select only one of these options, using the sum is the most reliable way to identify all such objects. We find that 1785 (or 8\%) of  the galaxies meet these criteria, and of these 445 (2\%) are found to have the largest vote for ``merger", 137 (0.6\%)  for ``disturbed" and 1203 (5.4\%) for ``irregular". As these are a small fraction of the sample removing them makes little difference to the results below, never-the-less we remove them in what follows and proceed with 20,254 ``normal" (or not ``odd") galaxies. 

We make use of Petrosian aperture photometry from SDSS in the $ugriz$ bands. These are k-corrected as described in \citet{Bamford2009}. Stellar masses are estimated from the colour-dependent mass-light ratio calibration presented by \citet{Baldry2008}. 

\section{Morphology of Local Galaxies} \label{demographics}

\begin{figure*}
\includegraphics[height=160mm,angle=-90]{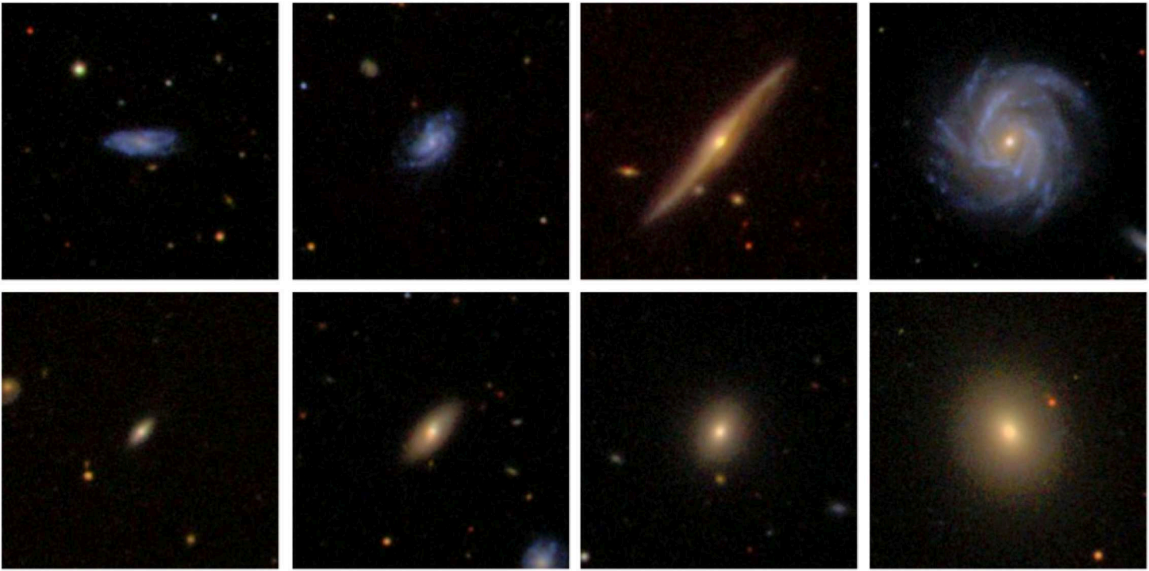}
\caption{Randomly selected example images of galaxies classified as either ``featured" (top row) or ``smooth" (bottom row) from Galaxy Zoo as a function of $r$-band absolute magnitude (brighter to the right).  All galaxies in this image are selected to have a redshift $z=0.03$, so are shown at the same physical resolution. Images are $gri$ composites from SDSS with a scale of 1.7 square arcmin.  \label{examples}}
\end{figure*}

 Many published works with Galaxy Zoo classifications use thresholds of $p_{\rm smooth}>0.8$ and $p_{\rm features}>0.8$ to identify samples of cleanly classified galaxies. With these cuts, we find that {28\%} of galaxies in the sample are clearly ``featured", and {24\%} are clearly ``smooth", (the remaining 48\% have only lower consensus classifications; this can include genuinely intermediate type galaxies, but also any galaxy where volunteers did not have clear consensus on morphology for reasons to do with the imaging rather than the galaxy itself). While galaxies with $p_{\rm smooth}$ and $p_{\rm features}<0.8$ are sometimes described as ``uncertain" and removed from studies \citep[e.g.][]{Schawinski2014}, information is contained in the lower agreement classifications. Relaxing the thresholds to use the majority answer for all galaxies in the sample allows every galaxy to be put into some category, although with increased uncertainty near the threshold. With this cut, which is similar, but not identical to  $p_{\rm smooth}>0.5$ or $p_{\rm features}>0.5$, as well as the vote fraction thresholds for classification recommended in Table 3 of W13, we find {50\%} of the normal galaxies in our volume limited sample to $z<0.035$ are best identified as ``featured" (mostly spirals, but barred and/or edge-on S0s and non-spiral galaxies with significant dust-lanes would also likely be in the category), and {50\%} as ``smooth" (or ``early-type", meaning E and S0s seen face-on\footnote{We note that the issue of face-on and edge-on S0s finding themselves in different categories is not unique to Galaxy Zoo classifications (see the discussion in \citealt{Bamford2009}). Edge-on S0s are almost impossible to distinguish from edge-on spirals, and face-on S0s can best be distinguished from ellipticals by their light profile shape; which is hard to judge by eye.}). Random examples of these two classes at $z=0.03$ (the median redshift of the sample) and as a function of absolute magnitude are shown in Figure \ref{examples}. 

\subsection{Spiral Arms, Bars and Bulges}
 
  It is only possible to identify spiral arms, bars and other disc features in disc galaxies which are sufficiently face-on for these to be visible, so we want to exclude almost edge-on disc galaxies from our sample.  Among the galaxies identified as ``featured" in our ``normal" galaxy sample, we find {17\% ($N=1 699$)} have values of $p_{\rm edge-on}>0.8$. This is reassuringly close to the number of galaxies expected to be found within $10^\circ$ of inclination, $i=90^\circ$ in a randomly orientated sample of disc shaped objects. 
Conversely, W13 publish a recommended threshold for ``oblique" galaxies in which we can reliably identify disc features (e.g. bars, spirals) of $p_{\rm not~edge-on}>0.715$ (and $N_{\rm not~edge-on}>20$). In the sample discussed in this article, we find that {66}\% of the ``featured" galaxies fall into this group {($N=6 614$)}. 
 
 Of these ``oblique featured" galaxies: 
\begin{itemize}
\item {86}\% show clear spiral arms ($p_{\rm spiral} > 0.5$). Just {5\%} are found to have a vote fraction that strongly indicates the absence of spiral arms (\ie ~have $p_{\rm spiral}<0.2$). 
\item {31}\% have obvious bars ($p_{\rm bar}>0.5$). This strong bar fraction is consistent with previous Galaxy Zoo based work \citep[e.g.][]{Masters2011, Masters2012}. Weaker bars can be identified by $0.2<p_{\rm bar}<0.5$ \citep[e.g.][]{Skibba2012,Willett2013,Kruk2018}. Another {25}\% of the oblique spirals have weak bars by this definition, leaving just over {44\%} of oblique spirals without any clear sign of a bar feature (\ie~ $p_{\rm no~bar}>0.8$) at the scales detectable in the SDSS images (\ie ~1--2 kpc at these distances). 
\end{itemize}

Bars in GZ2 have been studied in many papers \citep[e.g.][]{Masters2011, Masters2012, Skibba2012, Cheung2013, Cheung2015, Galloway2015, Kruk2017, Kruk2018}, and the number of spiral arms have been investigated by \citet{Willett2015}, \citet{Hart2016} and \citet{Hart2017a}. \citet{Hart2017b,Hart2018} make use of automated pitch angle measures along with spiral arm numbers from Galaxy Zoo to investigate spiral arm formation mechanisms. However, this is the first paper to attempt to make use of the crowdsourced arm winding measures directly, so we will start by comparing them with the automated measures. 

\subsubsection{Crowdsourced Arm Winding and Bulge Size}

\begin{figure*}
\includegraphics[height=165mm,angle=-90]{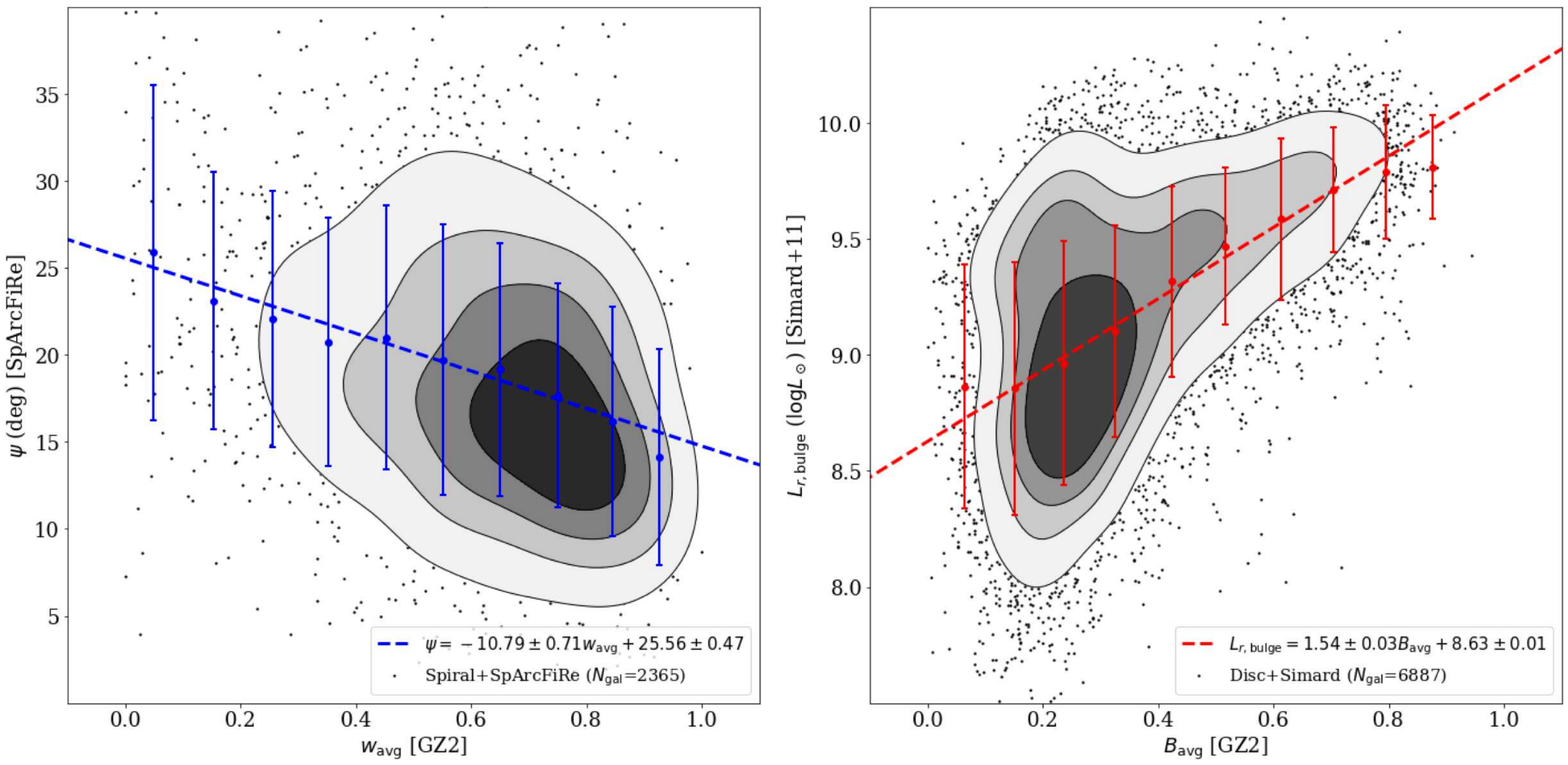}
\caption{(a) Galaxy Zoo winding score from Eq. \ref{eqn:winding} vs. measured pitch angles from SpArcFiRe for all spirals with at least one reliably identified arc ($N=2365$; see \citealt{DavisHayes2014} and \citealt{Hart2017b}). (b) Galaxy Zoo bulge prominence from Eq. \ref{eqn:bulge} vs. SDSS $r$-band bulge luminosity as measured from \citet{Simard2011} for a sample of $N=6887$ galaxies with both measures (see  \citealt{Hart2017b} for details of the sample selection). The dashed lines show the best fit straight line for each plot. The points with error bars show binned means plus/minus the scatter in each bin. \label{pitch}}
\end{figure*}

We define an arm winding score from Galaxy Zoo classifications as
\be
\label{eqn:winding}
w_{\rm avg} =  0.5p_{\rm medium} + 1.0p_{\rm tight}.
\ee
The choice of coefficients applied to these vote fractions is arbitrary, however this measure has the advantage of providing a single number measuring the tightness of the spiral arms as seen by Galaxy Zoo users, and will be $w_{\rm avg} = 1.0$ where the arms are most tightly wound and $w_{\rm avg} = 0.0$ where they are very loose. We compare these estimates with pitch angles measured by the SpArcFiRe method \citep{DavisHayes2014} in Figure \ref{pitch} (see \citealt{Hart2017b} for more details). This demonstrates how well arm winding as identified by Galaxy Zoo users correlates with pitch angle for those galaxies where pitch angle can be measured. The best fit trend gives 
\be
\label{eqn:pitch}
\Psi =  (25.6\pm0.5)^\circ - (10.8\pm0.8)^\circ w_{\rm avg}, 
\ee
where $\Psi$ is the pitch angle in degrees. This provides a way to estimate numerical pitch angles from the GZ2 visual descriptions, and reassurance that the crowdsources measures of arm winding are measuring a real property of the spiral arms. In the remainder of this article we will make use of $w_{\rm avg}$ from GZ2 directly.  

We define a bulge prominence from GZ2 using
\be
\label{eqn:bulge}
B_{\rm avg} = 0.2p_{\rm  just~noticeable} + 0.8p_{\rm obvious}+ 1.0p_{\rm dominant},
\ee
where $p_{\rm  just noticeable}$, $p_{\rm obvious}$ and $p_{\rm dominant}$ are the fractions of users who indicated the bulge was ``just noticeable", ``obvious" or ``domninant" respectively. This provides a single number, which ranges from $B_{\rm avg} = 0.0$ for galaxies with no bulge component, to $B_{\rm avg} = 1.0$ for spiral galaxies with dominant bulges, although we note there is no a priori reason to select these co-efficients specifically. On the right hand side of Figure \ref{pitch} we plot this measure of bulge prominence from GZ2 against the SDSS r-band luminosity of bulges as measured by \citet{Simard2011}. It is this quantity from \citet{Simard2011} which correlates most strongly with GZ2 bulge prominence (\ie ~not $B/T$).  The best fit trend is 
\be 
\log(L_{\rm r,bulge}/L_\odot) = 8.63\pm0.01 + (1.54\pm0.03) B_{\rm avg}.
\ee
There is significant scatter in this plot, particularly where the GZ2 classification indicates that bulges are not prominent. Some of the scatter will be caused by the use by \citet{Simard2011} of models which include only two components, a bulge and a disk, which is problematic when the galaxy has a strong bar. \citet{Kruk2018} demonstrated that in bulge+disc+bar decompositions, there was a stronger correlation of $B/T$ with Galaxy Zoo consensus classifications of bulge size. The \citet{Simard2011} decompositions also struggle when there is a significant dust lane. Even with this large scatter, as there is a positive correlation between an automated measure of bulge size and our GZ2 bulge prominence parameter, we will proceed to make use of the latter below. 
 
\subsection{The Correlation of Bulge Size and Spiral Arm Tightness}

As described in \S 1, the classic Hubble sequence for spiral galaxies implies that bulge size and spiral arm winding should be highly correlated. It has long been recognised that this correlation is not perfect \citep[e.g.][]{1970ApJ...160..811F, kennicutt1981} and this is supported by more recent studies \citep{Hart2017b,Hart2018}, though others do claim to see a trend \citep[][when disc gas mass is also considered]{Davis2015}. In this section we investigate how tightly correlated bulge size and spiral arm tightness are found to be for galaxies with visible spiral arms in the Galaxy Zoo sample making use of the unique value of bulge size and spiral arm tightness from the GZ2 classifications as defined in Equations \ref{eqn:winding} and \ref{eqn:bulge}. In this scheme, a ``classic'' Sa would have coefficients of 1.0 (representing both a large bulge and tight arms), and a ``classic'' Sc would have coefficients of 0.0 (representing a small bulge and loose arms). 

We plot these values for the subsample of Galaxy Zoo galaxies which have reliable classifications for both - i.e.. those galaxies with visible spiral arms. We select this sample (as advised by \citealt{Willett2013}) using cuts on the classification votes in answers earlier up the GZ2 tree, specifically $p_{\rm features} > 0.430$, $p_{\rm not~edge-on} > 0.715$, $p_{\rm visible~arms} > 0.619$, and in addition require the number of people answering the question about spiral arm windiness to be at least 20. This gives a sample of $N = 4830$ spiral galaxies in which we can ask how well bulge size correlates with spiral arm winding angles.
 
\begin{figure}
\includegraphics[width=80mm]{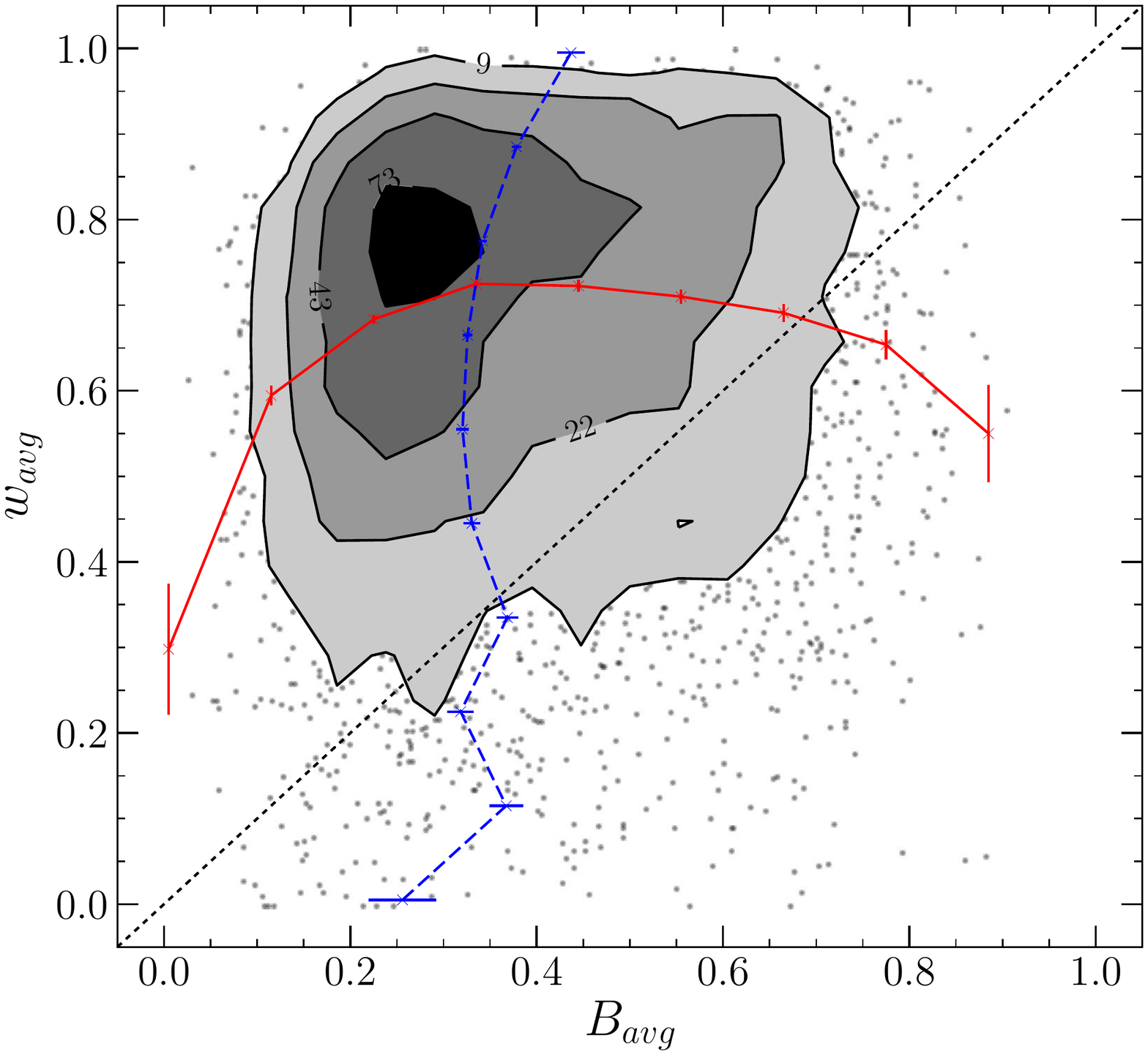}
\caption{We show here the location of {4830} nearby spiral galaxies on a plot of bulge size versus degree of arm winding as indicated by Galaxy Zoo classifications. The contours contain [0.5, 1, 1.5, 2] $\sigma$ of the 2D distribution in each plot (the numbers denote at least how many galaxies are contained in each bin enclosed by the contour). Points are shown at the lowest density.  The dotted line shows a 1-1 correlation between our two parameters; this is not necessarily what we expect for the classic spiral sequence but something with a upwards diagonal trend should be expected, with Sas at the upper right and Scs at lower left. This plot does not display that behaviour. The dashed lines show medians of bulge prominence (blue dashed line) and arm winding (red solid line) in bins of the respective other parameter.  \label{bulgewinding}}
\end{figure}

 We plot the measure of bulge size versus arm windiness for the oblique spiral sample in Figure \ref{bulgewinding}. In this sample of nearby almost face-on spiral galaxies we find no significant correlation between bulge size and arm windiness. There is a tendency for spirals with large bulges to have only tightly wound spirals (\ie ~both $w_{\rm avg}$ and $B_{\rm avg}$ are close to one), but for spirals with small bulges the complete range of spiral arm winding values are found almost as often as each other (although the median winding score still remains close to one, the distribution is rather flat). The median of the alternate value in bins of bulge prominence (red) and arm winding (blue) are shown, which highlight the lack of any strong trend. Despite the typical picture, this is consistent with the previous literature, in that Sa galaxies (as defined by arm winding) have been discussed with both large and small bulges \citep[e.g.][]{Hogg1993}, while Sc galaxies (as defined by loose arms) are only ever discussed with small bulges. This plot does not show the any clear sign of the diagonal trend implied by the strictest definition of the spiral sequence.

There is a possibility that pitch angles will be impacted by the inclination of the spiral galaxy. For a quantitative discussion of this effect see \citet{Block1999} who find it to only be important for highly inclined galaxies. These are excluded from our sample so we do not expect inclination to have a large effect, never-the-less we still check our result in bins of axial ratio, and find no significant changes in the observed pattern.

\begin{figure*}
\includegraphics[width=160mm,angle=0]{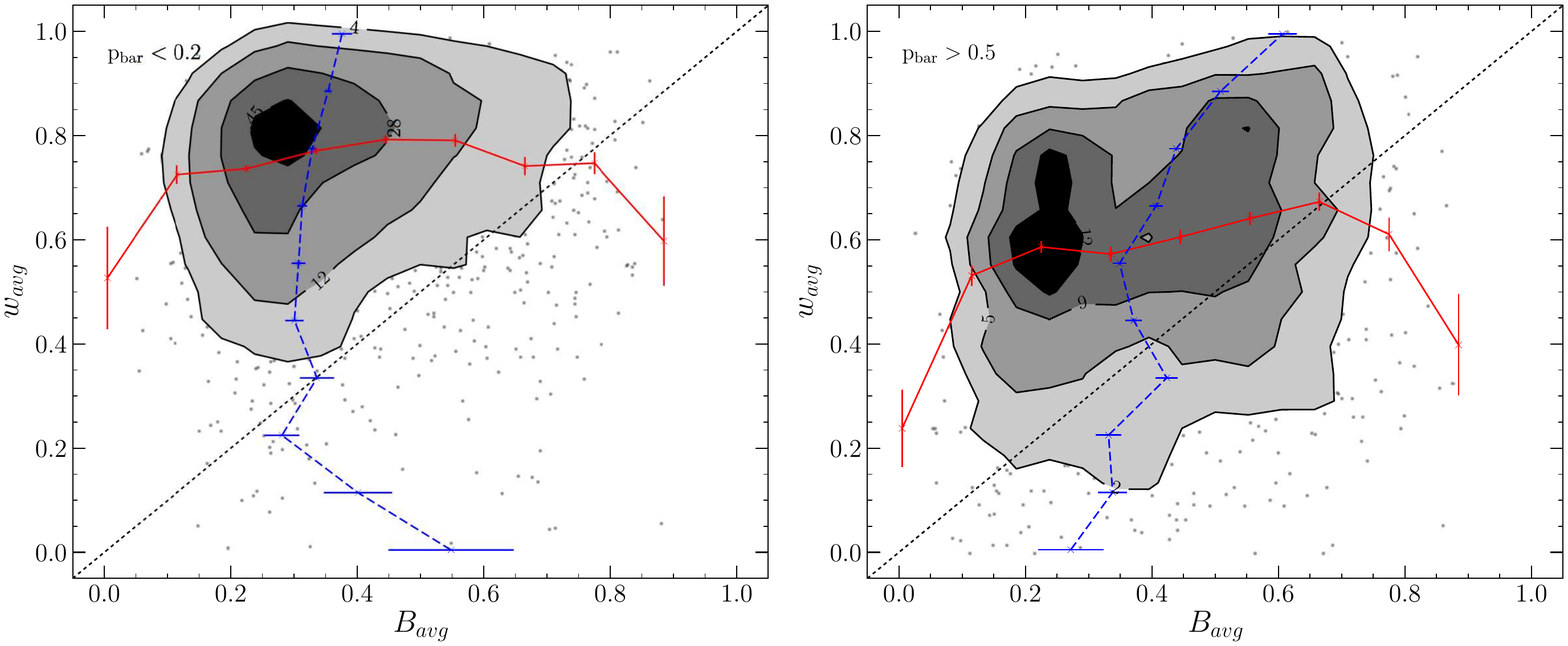}
\caption{As Figure \ref{bulgewinding} but for subsamples of the oblique spirals split by bar classification.  Left panel: galaxies with $p_{\rm bar} < 0.2$; right panel: galaxies with $p_{\rm bar} > 0.5$ \label{bars}. The classic spiral sequence is a diagonal line in this plot (not necessarily the dotted line of 1-1 trend which is shown) with Sas at the upper right and Scs at lower left. In neither sub-sample does the data display that behaviour, and it is particularly absent in the sub-sample of barred spirals (right). The dashed lines show medians of bulge prominence (blue dashed line) and arm winding (red solid line) in bins of the respective other parameter.  }
\end{figure*}

 Given the traditional Hubble tuning fork is split by bar classification, and also because the presence of a bar can confuse both automated and crowdsourced measures of both bulge size and spiral pitch angle, we also split the sample based on the presence or absence of a strong bar (as shown in Figure \ref{bars}). We find that spirals with strong bars ($p_{\rm bar}>0.5$ shown in the right panel of Figure \ref{bars}) were more likely to have larger bulges and less tightly wound spirals than those with no bars ($p_{\rm bar} < 0.2$ shown in the left panel of Figure \ref{bars}), and for a given bulge size, barred spirals will have looser arms than unbarred spirals, but there remains no clear correlation between bulge size and spiral arm pitch angle in either subgroup.  
  
\begin{figure*}
\center
\includegraphics[width=160mm]{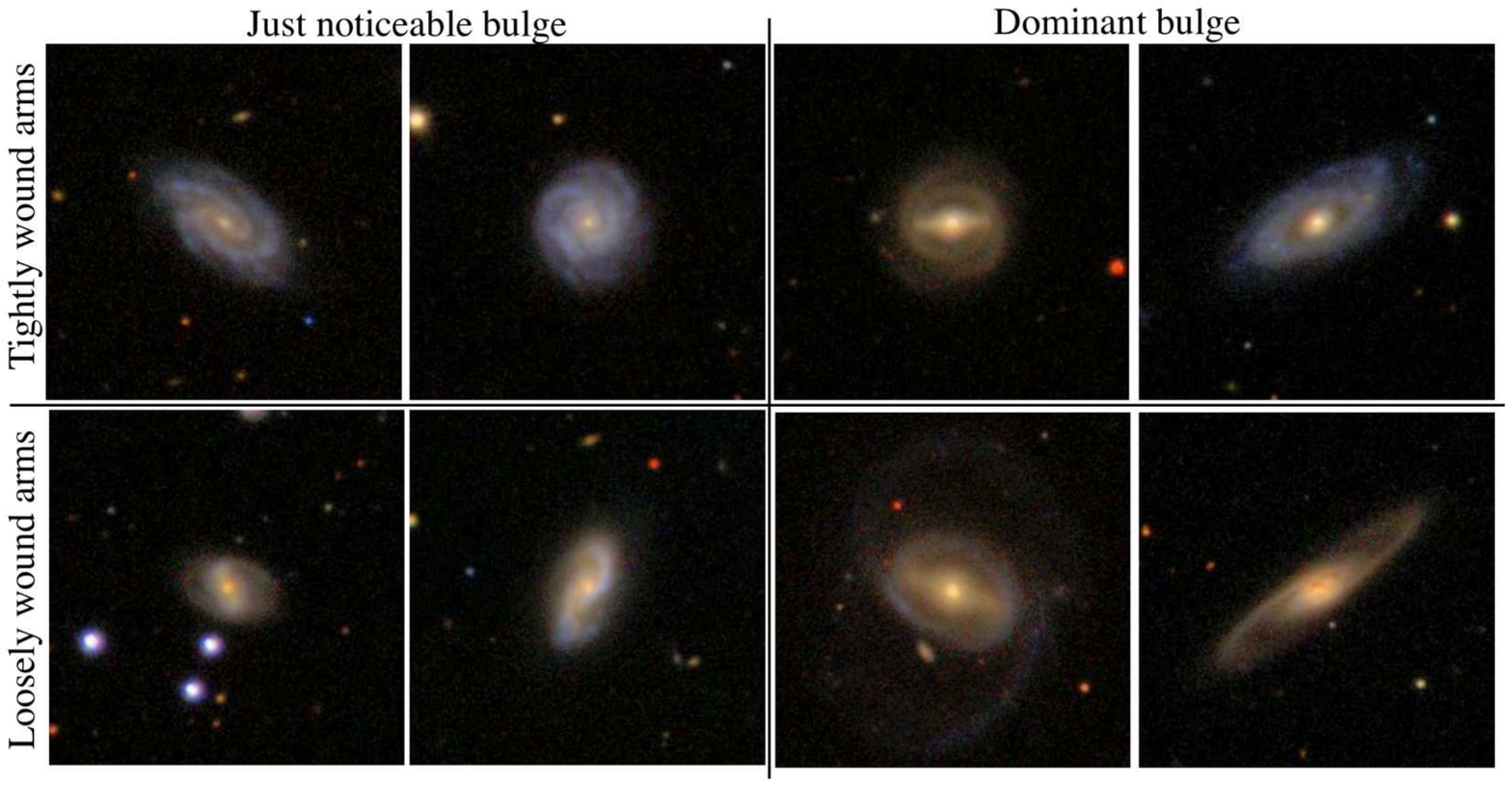}
\caption{Example images of galaxies at $z=0.03$ and $M_r\sim -21$ with both tightly wound and loose spiral arms (upper and lower rows respectively) and small or large bulges (left and right columns respectively). In each case galaxies are shown with either strong bars ($p_{\rm bar}>0.5$) or no bar ($p_{\rm bar} < 0.2$) to the right or left in each section. Images are $gri$ composites from SDSS with a scale of 1.7\arcmin~ square.  \label{windingexample}.}
\end{figure*}

 We show in Figure \ref{windingexample} examples of galaxies at $z=0.03$ from the four quadrants of Figure \ref{bulgewinding} (\ie ~the traditional Sa and Sc types, but also spirals with a small bulge and tightly wound arms, and those with large bulges and loosely wound arms) with either strong bars ($p_{\rm bar}>0.5$) or no bar ($p_{\rm bar} < 0.2$). 
 
\section{Discussion} \label{discussion}
 
 We have previously (in W13) discussed how best to assign $T$-types to Galaxy Zoo galaxies from the classification votes in GZ2. As is conventional, both the votes for tightness of spiral arms, and bulge size were considered. In that work, however, we concluded that modern expert visual classification of spiral Hubble types (based on comparison with both \citet[hereafter NA10]{Nair2010a} and \citet{EFIGI} was primarily driven by bulge size, independent of the tightness of spiral arms, with the best fitting relation (based on symbolic regression) being found to be
 \be
 T = 4.63 + 4.17~p_{\rm no~bulge} - 2.27~p_{\rm obvious} - 8.38~p_{\rm dominant}. 
 \ee 
 
\begin{figure}
\includegraphics[width=60mm,angle=-90]{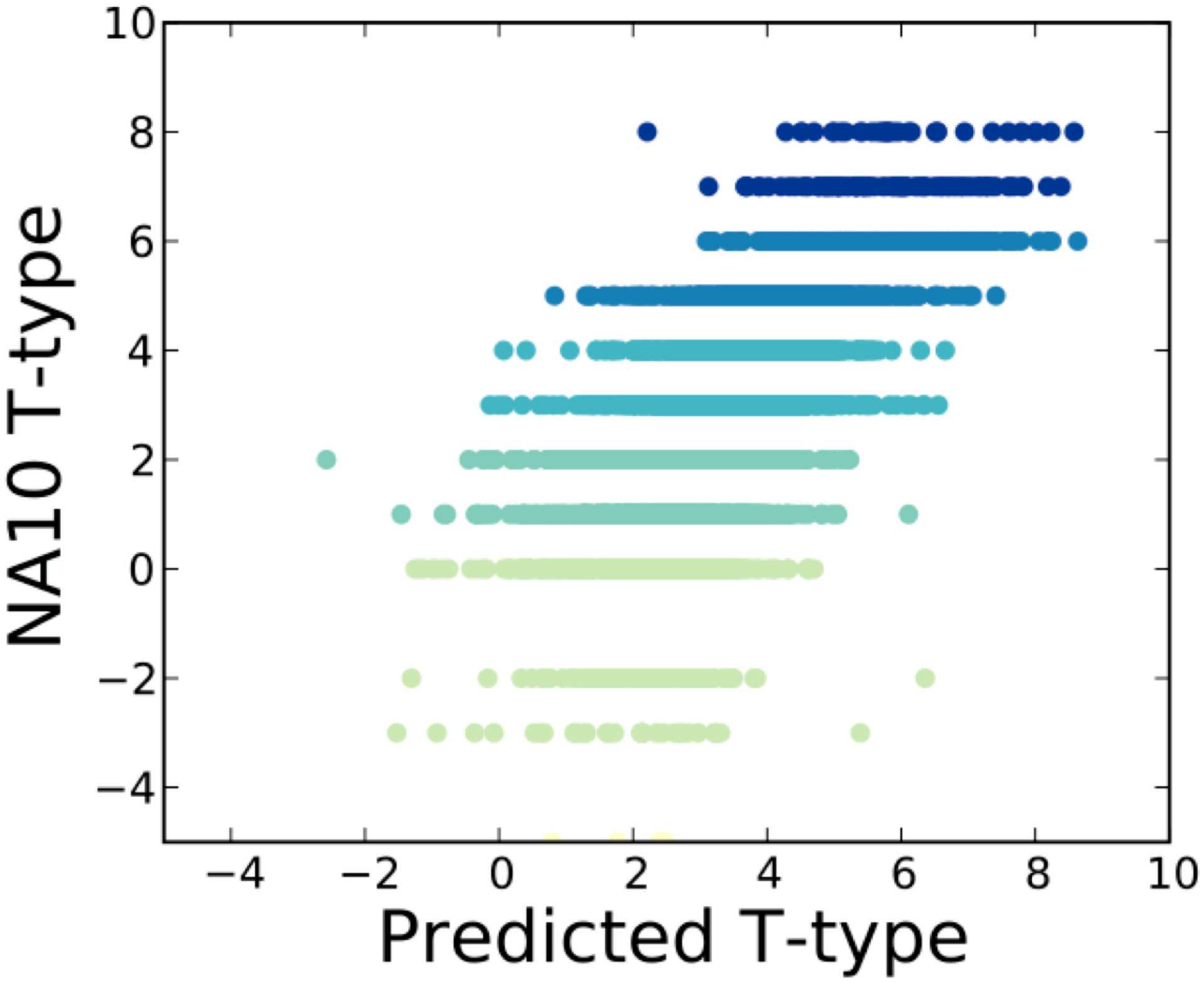}
\caption{Predicted T-type classifications as fit by \citet{Willett2013} for GZ2 galaxies shown versus their T-types from Nair \& Abraham 2010. Galaxies are colour coded by their morphologies as identified by NA10 (as indicated on the y-axis). Galaxies shown are only those with sufficient answers to characterise the arms winding and arms number GZ2 tasks, which selects heavily for late-type galaxies. This explains the lack of ellipticals in the plot, but highlights the fact that S0/a and earlier galaxies (Types $\leq 0$) do not agree well with the linear sequence. Reproduced from \citet{Willett2013}.  \label{T-type}}
\end{figure}

We point the interested reader to the lower panel of Figure 19 from W13 (reproduced for convenience in Figure \ref{T-type}) which compares the predicted T-types from the above equation to the T-types assigned by NA10. As was pointed out in W13, this work, along with other comparisons with recent expert visual classifications (e.g. the EFIGI sample of \citealt{EFIGI}, or the recent work of \citealt{Gao2019}) demonstrate clearly that the modern spiral Hubble sequence is defined by bulge size alone, with little reference to spiral arm tightness. We also draw the reader's attention to \citet{deJong1996} who show a correlation between bulge magnitude and Hubble type in their sample of 86 galaxies (albeit with large scatter, which they interpreted as meaning bulge size was a bad predictor for spiral Hubble type). These results appear to be in clear contrast to the result highlighted in \citet{kennicutt1981} who show that the \citet{sandagetammann1981} classifications of spiral type correlate best with pitch angle in their sample (although they do note significant scatter, which was used by \citet{deJong1996} to argue there was no tight correlation between pitch angle and Hubble type). Significant scatter/a lack of correlation between pitch angle and Hubble type has also previously been observed in a sample of 45 face-on spirals presented \citet{SeigarJames1998a}; although \citet{SeigarJames1998b} also see little correlation between Hubble type and bulge size in the same sample.

It is clear from all this that there has long been some uncertainty over the main driver of Hubble spiral type classification. It is perfectly normal for scientific classification schemes to change over time as more information becomes available (e.g. the reclassification of Pluto is the most discussed recent example in astronomy, \citealt{Messeri}), and there are many arguments that would have bulge prominence as the more fundamental property rather than spiral arm tightness. Bulge size is known to correlate with star formation properties \citep[e.g.][]{Cheung2012,Fang2013}, and morphological quenching (the growth of bulge component or the diminishment of the disc) has been suggested as a mechanism for the quenching of star formation \citep{Martig2009}. However we wish to clearly point out that because of this shift in definition, {\it traditional and modern definitions of spiral types (often with the same names used) do not map uniquely}, and therefore care should be taken when comparing results using different schemes.  

 Regardless of the description of the traditional spiral sequence classification, many models of spiral arm formation (see \citealt{DobbsBaba2014} for a recent and comprehensive review) do not predict that spiral arm pitch angle should correlate with bulge size. For example, in swing amplification models pitch angle should correlate best local shear in the disc, which is related most strongly to total galaxy mass and the shape of the rotation curve (e.g. see the simulations of \citet{Grand2013,Michikoshi2014}; this has also been observed in small samples of real galaxies by \citet{Seigar2005,Seigar2006}, however \citet{Yu2018,Yu2019} do not confirm this strong correlation, and furthermore \citet{Yu2019} suggest that this is because up to a third of the pitch angles used in the previous work are significantly overestimated). In these models pitch angle in turn correlates with spiral arm amplitude. Tidally induced spirals should have pitch angles which correlate with the strength of the interaction, rather than any internal properties of the galaxy \citep{Kendall2011}.  In the classic static density wave model of \citet{LinShu1964} however, there is expected to be a correlation between pitch angle and inner mass density, which was laid out in \citet{Roberts1975}. \citet{Bertin1989a,Bertin1989b} explore a modal density wave model and also find a correlation between pitch angle, interior mass (often interpreted as bulge mass) and gas density in the disc. It is therefore extremely interesting to consider how well pitch angle and bulge size correlate in larger samples.
 
 Our result, in Figure 5, shows only a loose correlation between the two parameters of the spiral sequence. Galaxy Zoo classifications of spiral arms have previously been studied by \citet{Hart2017b,Hart2018}, who also found a weak correlation between pitch angle and bulge size and used it to argue that most spiral arms in the local Universe (up to 60\%) are not caused by swing amplified density waves. The results presented here, also appear to be contrary to the predictions of the classic static density wave model. 

The static spiral density wave model \citep{LinShu1964} was originally conceived to solve the ``spiral winding problem" and create static, long lived spiral arms, thus explaining their ubiquity in the local Universe. As first discussed by \citet{Oort1962} as the ``winding dilemma", if spiral arms were material in nature, the differential rotation of galactic discs would rapidly wind them up and destroy them. Prior to the \citet{LinShu1964} model there was no known mechanism which could create long lived spirals; something of an embarrassment for a field in which the majority of objects studied showed spiral structure! However, it is notable that both observational data and models have slowly, but consistently moved the discussion of both extragalactic spiral arms \citep{2013ApJ...766...34D} and the spiral arms in our own Galaxy \citep{hunt2018} away from a view of static density waves, towards a variety of transient models, all of which allow winding in some form (e.g. see \citealt{Sellwood2011} who clearly lays out the evidence for transient spirals in external galaxies and our own Milky Way, and also \citealt{Merrifield2006} who discuss a short lived grand-design spiral). Indeed, in recent years there is a growing consensus that spiral arms must wind over time, which is supported by our observation of no strong correlation between bulge size and arm winding. This shift in the community has in part been driven by the growing sophistication of simulations of spiral structure \citep[e.g.][]{Grand2013,Grand2017,Forgan2018,PettittWadsley2018}, but also the fact that so few observations show the proper signatures for density waves (e.g. \citealt{Merrifield2006,hunt2018} and see \citealt{Sellwood2011} for a comprehensive review).

  As a concrete example, \citet{PettittWadsley2018} investigate the dependence of pattern speeds and wind-up rates on morphology in a sample of 5 model galaxies (designed to mimic M31, NGC4414, M33, M81 and the Milky Way). They were interested in investigating the impact of changing bar and disk properties, however, bulge mass also varies between their models, and there is a clear suggestion in their results that the wind-up rate is affected by bulge mass. Their model of M33, which has a bulge-to-disk mass ratio almost an order of magnitude lower than the other systems, has the slowest wind-up rate of all of their simulations. Unfortunately the number of galaxies was too small to see the impact of the bar independently of bulge size, but we note that their slowest winding model also hosted a strong bar, which is in agreement with our observations of looser arms in general in galaxies with strong bars.

If the rate of winding is dependent on the mass of the central concentration, then there is a natural explanation for the observation we present here. Systems with large bulges would, quickly (formally we mean quickly compared to the dynamical time), develop tighter spiral arms, leading to the absence of systems with large bulges and loose arms that we observe. We would, in this model, expect systems with smaller bulges to have a range of spiral arm types, just as observed. We therefore suggest that our observations could support the idea that the majority of spiral arm structure observed in the Universe is transient and winding, rather than static grand design density waves. However as a caveat we note that there could be other reasons for the observed scatter of pitch angle at a given bulge mass; one suggestion is disc gas densities (e.g. as used in \citealt{Davis2015} to identify their ``fundamental plane of spiral structure"); if galaxies with small bulges had a wider range of observed disc gas densities than those with small bulges that could also explain our observations. This would be an interesting topic for future study. 

If our speculation is correct that these data show that many spiral arms wind up, their prevalence clearly indicates continued triggering of new arms. This suggests that we should see looser spiral features coexisting with tighter ones in many individual galaxies. In a set of 3-armed spirals in Galaxy Zoo indeed it is  common to see one as the odd arm out in pitch angle (Colin Hancock priv. comm). Further investigation of this allowing spiral arm pitch angles to vary across a single galaxy (as a function of arm) may reveal interesting physics. 

 Bars have commonly been invoked as drivers of $m=2$ density waves in spirals \citep[e.g.][]{DobbsBaba2014,Hart2017b}. The fact that we (and \citealt{Hart2017b}) observe that galaxies with strong bars have looser arms for the same bulge size supports this idea. This tells us that either the bar acts to slow down arm winding, or that it drives the $m=2$ mode such that the spiral arms do not wind. \citet{Hart2017b} discuss the role of bars in driving $m=2$ armed spirals in some detail. In that work, they quantify the effect of observing looser arms in barred spirals -- measuring pitch angles for spirals with bars as $4-6^\circ$ looser than in similar unbarred spirals. There are models for spiral arm formation which predict looser arms when driven by strong bars (the invariant manifold theory of \citealt{Romero-Gomez2007}). We also note that both loose spiral arms and bars are known to correlate with increased local density \citep{Casteels2013}, suggesting that in some galaxies they are both triggered in galaxy interactions. Spirals arms in barred galaxies may therefore be quite different in character to those in their unbarred cousins. 
 
 Our results (and those of \citealt{Hart2017a,Hart2017b,Hart2018}) are significantly different to those presented by \citet{Davis2015,Davis2017} who observe a strong correlation between bulge mass, gas density and pitch angle in their samples of spiral galaxies. In \citet{Davis2015} they measure pitch angle for a subset of 24 galaxies in the DiskMass PPAK sample \citep{Bershady2010,Martinsson2013}, a set of nearby nearly face-on spirals which was selected to be regular, not have large bars or bulges, or significant spiral perturbations and no strong kinematic disturbances (e.g. the streaming motions which might be associated with strong spiral arms), while in \citet{Davis2017} the sample is 44 very nearby spiral galaxies with measurements of central supermassive black hole masses (they note that most of these galaxies do have bars). It is curious that such a strong correlation was observed in these small samples, while we see very weak correlation in our volume limited sample of almost 5000 galaxies, so we explore the correlation using a simple selection criteria looking at a subsample of only the brightest galaxies in our sample (i.e. applying an apparent magnitude limit), and find that a correlation between bulge size and arm winding, although still with large scatter, is apparent in the brightest 100 galaxies, a trend largely driven by a lack of spirals with small bulge and tightly wound arms in this subset. Given this, it appears that the correlation seen by both \citet{Hubble1926} and most recently by \citet{Davis2015,Davis2017} may be a result of the implicit apparent magnitude limit on the sample selection, and disappears in a volume limited sample. 
  
  Given our results we caution against the use of pitch angle to estimate black hole mass which based on our results, we believe is unlikely to provide reliable black hole masses for large samples of spiral galaxies. The suggestion of a correlation between spiral arm pitch angle and supermassive black hole mass was first noted in a sample of 27 galaxies by \citet{Seigar2008}, and followed up with a larger sample (of 34 galaxies) by \citet{Berrier2013}, and 44 galaxies in \citet{Davis2017}. We are unsure what fraction of these galaxies may suffer from the significant pitch angle measurement errors noted by \citet{Yu2019} in papers from the same group \citep{Seigar2005,Seigar2006}, however given the clear uncertainty and confusion in the literature over not only the formation mechanism for spiral arms, but also which galaxy properties observed pitch angles best correlate with, we at best recommend caution in using pitch angles to estimate black holes masses. It is interesting that \citet{MutluPakdil2018} find a correlation between pitch angle and black hole mass in a random sample of 95 galaxies from the Ilustris simulation \citet{Illustris}, however they note the correlation between pitch angle and halo mass is stronger. Even if there is a correlation between pitch angle and supermassive black hole mass, this would not necessarily disagree with the lack of correlation between pitch angle and bulge mass observed here, if galaxy mass, rather than bulge mass were the main driver of the $M_\bullet$--$M_{\rm b}$ relation, as suggested by both observations \citet{Simmons2013,Simmons2017} in samples of bulge-less spirals with black hole mass estimates, and also in simulations \citep{Martin2018}.

\section{Summary}

We present the morphological demographics of a sample of bright ($M_r <-19$), nearby ($0.01<z<0.035$) galaxies with classifications from the Galaxy Zoo project. We find that {92\%} of these galaxies show the ``normal" morphologies found on the classic Hubble sequence, with just {8\%} classified as irregular, disturbed or merging. 

Among the ``normal" galaxies we find that in a nearby volume limited sample ($z<0.035$), ``featured" galaxies (which are overwhelmingly spiral galaxies) make up ~50\% of the sample. In this selection, we find that the fraction of edge-on spirals is as expected for a sample of randomly orientated discs, and define a sample of ``oblique" spirals which are face-on enough for disc features to be identified. 

 Among these ``oblique spirals" we find that {31\%} have strong bars, and {44\%} have no bars (\ie ~up to 56\% are consistent with having a bar of some kind) . The majority have clearly identified spirals ({86\%}), with just {5\%} having a consensus vote indicating a lack of spiral arms. These are likely S0 type galaxies with rings or bars \footnote{We note that S0 galaxies without any features are likely to be found in the ``smooth" arm of Galaxy Zoo classifications, and may be best identified via their aximuthally averaged light profile shape.}. 
 
 We use this sample to demonstrate that modern expert visual classification has moved away from the classic ``Hubble sequence" which prioritised spiral arm winding type over bulge size (e.g. allowing for small bulged Sa galaxies) and is now predominately a sequence ordered on central bulge size. This was previously noted by \citet{Willett2013}. Authors who make use of morphologies, particularly those drawn from different classifications, should take care that they understand well what is driving their morphological classifications; our results suggest that traditional morphologies (e.g. the RC3 \citep{1991rc3..book.....D}, or those found in the NASA/IPAC Extragalactic Database; NED\footnote{\tt http://ned.ipac.caltech.edu/}) do not map well onto current bulge-size based classifications. We also note that classifications based on bulge prominence will be more vulnerable to morphological k-correction 
than those using arm geometry. This seems particularly ironic to note at a time when we have large volumes of data where we would like to reduce the effect of redshift variation of classifications. 
 
Among the spiral galaxies, we find little or no correlation between spiral arm winding tightness and bulge size. Although spirals with large bulges are found to typically have tightly wound  arms, those with small bulges are found with a wide range of spiral arm pitch angles. We discuss how this could be interpreted as favouring winding models of spiral arms, with the winding rate dependent on the bulge size. There may be no ``winding problem" for spirals after-all, but rather spiral arms are constantly reforming. This predicts spirals ought to be found with a variety of pitch angles in a single galaxy, which should be tested; we also encourage further investigation into the source of the scatter of pitch angles in disc galaxies with different bulge sizes, which may reveal alternative explanations. 

Finally, we find that the presence of a strong bar tends to correspond to more loosely wound arms and larger bulges. This could be used to suggest that the presence of a strong bar in a galactic disc either prevents winding, or perhaps even drives static density wave spirals. 

 New higher resolution and deeper imaging of significant fractions of the sky from surveys like LSST and Euclid will provide significantly more galaxies with well resolved internal structure in the near future. Furthermore large scale integral field unit (IFU) surveys like SAMI \citep{Bryant2015} and MaNGA \citep{Bundy2015} are revealing how well morphology correlates with the underlying dynamics of the stars \citep{Foster2018}. These developments make galaxy morphological classification as relevant today to our understanding of galaxy formation and evolution as it was in the time of \citet{Hubble1926}, so we should take care to be precise about what we mean by morphological types. 
  
\paragraph*{ACKNOWLEDGEMENTS.} 

This publication has been made possible by the participation of more than 200,000 volunteers in the Galaxy Zoo project. Their contributions are individually acknowledged at \texttt{http://authors.galaxyzoo.org}.  We particularly wish to acknowledge the contributions of volunteers who participated in the Galaxy Zoo literature search activity (described at {\tt http://blog.galaxyzoo.org/2017/09/28/ galaxy-zoo-literature-search/}). These volunteers helped the authors identify literature examples of certain use of classification systems of galaxies. Finally we acknowledge the numerous contributors to the Galaxy Zoo Forum NGC Catalogue List ({\tt http://www.galaxyzooforum.org/index.php?topic=280028.0}) who made finding SDSS images of NGC galaxies so easy. Galaxy Zoo 2 was developed with the help of a grant from The Leverhulme Trust.

We thank Hugh Dickinson of the Galaxy Zoo Science team for helpful comments on an advanced draft, and Kyle Willett (formerly of the Galaxy Zoo Science team) for significant contributions in the early stages of this work. 

Funding for the SDSS and SDSS-II has been provided by the Alfred P. Sloan Foundation, the Participating Institutions, the National Science Foundation, the U.S. Department of Energy, the National Aeronautics and Space Administration, the Japanese Monbukagakusho, the Max Planck Society, and the Higher Education Funding Council for England. The SDSS Web Site is http://www.sdss.org/. 

The SDSS is managed by the Astrophysical Research Consortium for the Participating Institutions. The Participating Institutions are the American Museum of Natural History, Astrophysical  Institute Potsdam, University of Basel, University of Cambridge, 
Case Western Reserve University, University of Chicago, Drexel University, Fermilab, the Institute for Advanced Study, the Japan 
Participation Group, Johns Hopkins University, the Joint Institute for Nuclear Astrophysics, the Kavli Institute for Particle Astrophysics and Cosmology, the Korean Scientist Group, the Chinese Academy of Sciences (LAMOST), Los Alamos National Laboratory, the Max-Planck-Institute for Astronomy (MPIA), the Max-Planck-Institute for Astrophysics (MPA), New Mexico State University, Ohio State University, University of Pittsburgh, University of Portsmouth, Princeton University, the United States Naval Observatory and the University of Washington. 

This research made use of \textsc{astropy}, a community-developed core \textsc{python} ({\tt http://www.python.org}) package for Astronomy \citep{2013A&A...558A..33A}; \textsc{ipython} \citep{PER-GRA:2007}; \textsc{matplotlib} \citep{Hunter:2007}; \textsc{numpy} \citep{:/content/aip/journal/cise/13/2/10.1109/MCSE.2011.37}; \textsc{scipy} \citep{citescipy}; and \textsc{topcat} \citep{2005ASPC..347...29T}.

REH and SJK acknowledge funding from the Science and Technology Facilities Council (STFC) Under a Studentship and Grant Code ST/MJ0371X/1 respectively. RJS gratefully acknowledges funding from the Ogden Trust. BDS acknowledges support from the National Aeronautics and Space Administration (NASA) through Einstein Postdoctoral Fellowship Award Number PF5-160143 issued by the Chandra X-ray Observatory Center, which is operated by the Smithsonian Astrophysical Observatory for and on behalf of NASA under contract NAS8-03060.

\bibliography{gzreferences}{}
\bibliographystyle{apj}

\end{document}